\documentclass[twoside,a4paper,11pt]{proceedings}
\usepackage{graphicx}
\usepackage[hidelinks]{hyperref}
\usepackage{natbib}

\usepackage[utf8]{inputenc}  			
\usepackage[T1]{fontenc}				

\begin{document}
\pagenumbering{arabic}
\pagestyle{myheadings}
\thispagestyle{empty}
\vspace*{-1cm}
{\flushleft\includegraphics[width=3cm,viewport=0 -30 200 -20]{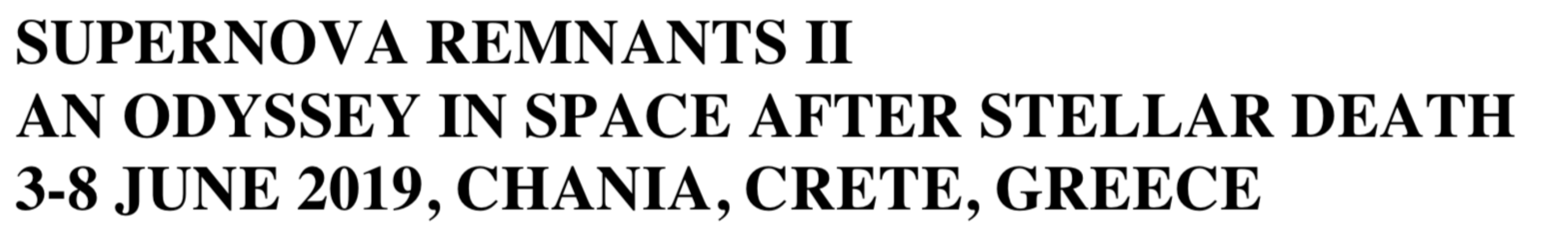}}
\vspace*{0.2cm}
\begin{flushleft}
{\bf {\LARGE
O\,CAESAR: The Optical CAtalogue of Extragalactic SupernovA Remnants
}\\
\vspace*{0.9cm}
I. Moumen$^{1,2}$,
C. Robert$^{1}$, 
D. Devost$^{2}$, 
L. Rousseau-Nepton$^{2,3}$,
D. Patnaude$^{4}$, \\
S. Safi-Harb$^{5}$,
R. P. Martin$^{3}$, 
L. Drissen$^{1}$, 
and T. Martin$^{1}$ \\
%
}
\vspace*{0.3cm}
%
$^{1}$ \small{Département de physique, de génie physique et d’optique, Université Laval, and Centre de \\ 
 Recherche en Astrophysique du Québec (CRAQ), Québec, QC, G1V 0A6, Canada} \\
$^{2}$ \small{Canada-France-Hawaii Telescope, Kamuela, HI, 96743, USA} \\
$^{3}$ \small{Department of Physics and Astronomy, University of Hawaii at Hilo, Hilo, HI, 96720, USA} \\
$^{4}$ \small{Smithsonian Astrophysical Observatory, Cambridge, MA 02138, USA} \\
$^{5}$ \small{Department of Physics \& Astronomy, University of Manitoba, Winnipeg, MB R3T 2N2, Canada}
%
\end{flushleft}
\markboth{
O\,CAESAR: The Optical CAtalogue of Extragalactic
SupernovA Remnants
}{
Moumen et al.
}
\thispagestyle{empty}
\vspace*{-0.3cm}
\begin{minipage}[l]{0.09\textwidth}
\ 
\end{minipage}
\begin{minipage}[r]{0.9\textwidth}
\vspace{1cm}
\section*{Abstract}{
\small{
We present O\,CAESAR, the Optical CAtalogue of Extragalactic SupernovA Remnants. 
O\,CAESAR will provide the largest homogeneous optical survey of extragalactic SNR candidates taken by the same telescope, the same instrument, and under similar observational conditions. 
Our sample is volume-limited (D $\leq$ 10 Mpc) and includes mostly galaxies ($\sim$40) from the Large Program  SIGNALS.
The Observations are carried out using SITELLE, the imaging Fourier transform spectrograph of the Canada-France-Hawaii Telescope. Using three filters, we are able to measure the strong emission lines [O{\,II}]$\lambda$3727, H$\beta$, [O\,III]$\lambda\lambda$4959,5007, H$\alpha$, [N\,II]$\lambda\lambda$6548,6583, and [S\,II]$\lambda\lambda$6716,6731. 
Identification of the SNR candidates will be done automatically and will be based on four criteria for regions where the emission lines flux ratio [S{\,II}]/H$\alpha$\,$\ge$\,0.4. 
To confirm the shock-heated nature of the ionization mechanism in the candidates derived from our sample, we adopted a self-consistent spectroscopic analysis using the whole set of emission lines available with our SITELLE data and considering Sabbadin plots and BPT diagrams.
We here present our method and  first results for the spiral galaxy NGC\,3344.
}
\vspace*{-0.3cm}

\normalsize}
\end{minipage}

\section{O CAESAR}
\vspace*{-0.4cm}
\nocite*{}
The goal of the new catalogue O\,CAESAR is to present the primary characteristics of a large sample of SNR candidates (i.e. coordinates of the emission peak, optical size, [S\,II]/H$\alpha$ ratio, strong emission line flux, and spectroscopic confirmation using Sabbadin plots$^{1}${\let\thefootnote\relax\footnote{{$^{1}$ Sabbadin plots (Sabbadin et al., 1977) and BPT diagrams (Baldwin et al. 1981)  are two sets of diagrams used to identify the ionization mechanism of the nebular gas based on the optical emission lines (H$\beta$, [O\,III]$\lambda$5007, [N\,II]$\lambda$6548, H$\alpha$, [N\,II]$\lambda$6583, [S\,II]$\lambda\lambda$6716,6731, and [O{\,I}]$\lambda$6300 when available).}}} and BPT diagrams$^{1}$; 
along with parameters related to the galaxy host (morphology, mass, metallicity, SFR, location within the galaxy structures).
This database will be homogeneous, as it will be extracted from the same instrument SITELLE (Spectromètre Imageur à Transformée de Fourier pour l'Etude en Long et en Large de raies d'Emission) at the Canada-France-Hawaii Telescope  (CFHT) with the same technical constraints and similar observing conditions. Finally, the SNR confirmation using other wavelength ranges (near-IR, radio, and X-rays) will be highlighted with references within the catalogue when possible. 
\vspace*{-0.1cm}

\section{The Sample, Observation, and Data reduction}
\vspace*{-0.3cm}
O CEASAR will be volume-limited (D\,$\leq$\,10\,Mpc) and will include all the galaxies from the CFHT Large Program SIGNALS  (Star formation, Ionized Gas and Nebular Abundances Legacy Survey with SITELLE;
Rousseau-Nepton et al. 2019;
Fig.~1a). 
Other galaxies may also be considered to provide a diversify of SNR environments.
The sample will be observed using the imaging Fourier transform spectrometer SITELLE (Drissen et al. 2019) installed on the 3.6-m CFHT, which offers a large field of view (11$^{\prime}$$\times$11$^{\prime}$), complete spatial coverage, and a high spatial resolution (0.32$^{\prime\prime}$  seeing-limited), making our observations ideal for covering the whole disk of nearby galaxies. 
For each galaxy, more than 4 million spectra will be obtained using the filters SN1 (365-385\,nm, R\,$\simeq$\,1800), SN2 (480-520\,nm, R\,$\simeq$\,2000), and SN3 (651-685\,nm, R\,$\simeq$\,5000) with a typical (seeing-limited) spatial resolution of 0.8$^{\prime\prime}$. 
Only $\sim$\,3 hours are needed to reach the spectral resolution requested for each filter for a total of 10 hours of integration time for each galaxy. 
Data reduction is performed with ORBS (Outils de Reduction Binoculaire pour SITELLE) and lines are fitted using ORCS (Outils de Réduction de Cubes Spectraux), two software tools developed specifically for SITELLE  (Martin\,et\,al.\,2016).
\vspace*{-0.6cm}

\section{SNR Identification and Confirmation}
\vspace*{-0.4cm}
In order to identify the SNR candidates in an objective way, we use a technique similar to the one described by Rousseau-Nepton et al. (2018). 
This technique was initially created to study the star-forming regions in NGC\,628.
Moumen et al. (2019) have adapted this technique to  automatically identify SNR candidates in the nearby galaxy NGC\,3344$^{2}${\let\thefootnote\relax\footnote{{$^{2}$ See also the contribution '3D Optical Spectroscopic Study of NGC3344 with
SITELLE: I. Identification and Confirmation of SNRs' in these proceedings.}}} (Fig.~1b). 
This work revealed more than 2000 emission regions (SNR, HII regions, and diffuse ionized gas regions). Next, four criteria were applied to select SNR candidates: 
    (i) Line ratio [S\,II]/H$\alpha$\,$\geq$\,0.4 (e.g. Fig.~1c); 
    (ii) A signal-to-noise ratio\,$\geq$\,5 for H$\alpha$ and [S\,II] lines; 
    (iii) A size of the emission region\,$\leq$\,120\,pc; 
     (iv) A correlation coefficient for the region H$\alpha$ flux profile\,$\geq$\,0.5 (Moumen\,et\,al.\,2019).
     This revealed 129 SNR candidates.
     SITELLE data provide emission lines (Fig.~1d) that are very important to classify the selected candidates (Fig.~1e) and to get their gas physical parameters, e.g. [S{\,II}]/H$\alpha$ and [O{\,II}] for the main shock heating mechanism, [N{\,II}]/H$\alpha$  and [O{\,II}]  for the metallicity, [O{\,III}]/H$\beta$ for the shock velocity, [S{\,II}] ratio for the density, H$\alpha$/H$\beta$ for the extinction, etc.). In order to confirm the shock-heated nature of the SNR candidates, we adopt a self-consistent analysis (Fig.~1f and 1g) using diagnostics from Sabbadin et al. (1977) and Baldwin et al. (1981). 
\vspace*{-0.4cm}

\section{What's next?}
\vspace*{-0.4cm}
Understanding emission nebulae like SNRs in various galaxies is a major issue in astrophysics that involves a statistical approach for large samples in galactic environments of all kinds. 
Instruments like SITELLE at the CFHT have opened a new era in the three-dimensional optical study of Galactic and extragalactic emission nebulae. 
In this work, we  presented O\,CAESAR that will provide the largest homogeneous optical survey of SNR candidates in nearby galaxies avoiding the main issues of the study of Galactic SNRs.
For each galaxy, all the SNR candidates are at the same distance and are not affected by
the high absorption in the Galactic plane. 
Furthermore, O\,CAESAR will provide synergies with works at other wavelengths in nearby galaxies.
Finaly, various sets of model of photo-ionization and shocks make it possible to connect these types of observations to the physical conditions (temperature, density, chemical composition, ionizing source) in the nebulae. But more sophisticated models including a large set of parameters are still needed. O\,CAESAR$^{3}${\let\thefootnote\relax\footnote{{$^{3}$ O CAESAR's website: \url{http://www.ismael.free.fr/o-caesar}}}} will be an excellent source of spectroscopic data for developing and testing new photo-ionization and shock models. 
\vspace*{-0.4cm}

\begin{figure}
\center
\includegraphics[width=\textwidth]{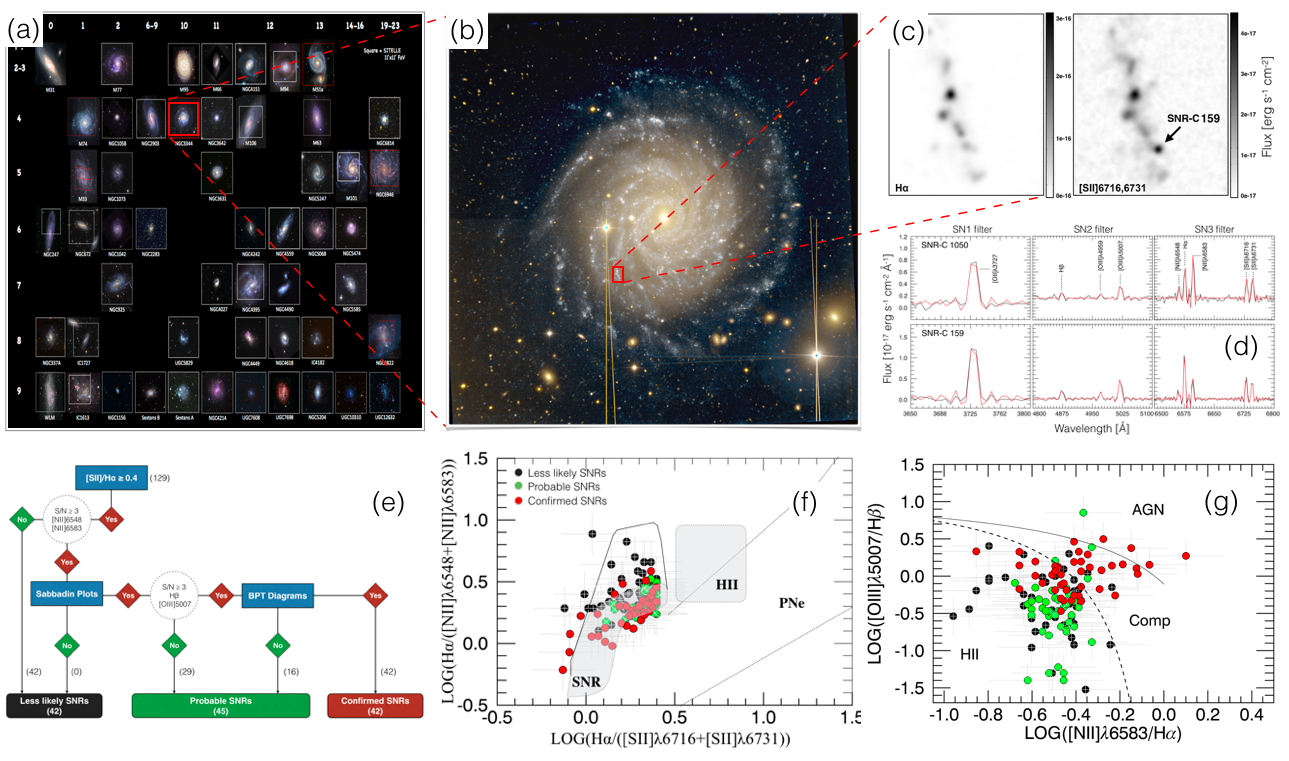} 
\caption{
\footnotesize{(a) SIGNALS’ galaxies  (Rousseau-Nepton\,et\,al.\,2019); 
(b) The SITELLE deep image of NGC\,3344 obtained by summing the interferograms of the SN1 (blue) and SN3 (red) filters (Moumen\,et\,al.\,2019);
(c) H$\alpha$ and [S{\,II}]$\lambda\lambda$6716,6731 flux maps near of the candidate SNR-C159; 
(d) SITELLE spectra for one pixel within SNR-C159 and SNR-C1050 (in black, the observed spectrum and in red, the fit obtained with ORCS); 
(e) Scheme of our spectroscopic analysis used to classify the 129 SNR candidates found in NGC\,3344: it revealed 42 Confirmed SNRs, 45 Probable SNRs, and 42 Less likely SNRs; Exemples of 
(f) a Sabbadin plot and 
(g) a BPT diagram with the SNR candidates. (Note: All the panels are available in the online ressources.)}}
\vspace*{-0.4cm}
\end{figure}

%
%


\section*{References}
\vspace*{-0.4cm}
\footnotesize{Baldwin J.A., Phillips M.M., Terlevich R., 1981, PASP, 93, 5} || 
\footnotesize{Drissen L., et al., 2019, MNRAS, 485, 3930} 
\footnotesize{Martin T.B., Prunet S., Drissen L., 2016, MNRAS, 463, 4223} ||
\footnotesize{Moumen, I., et al., 2019, MNRAS, 488, 803} 
\footnotesize{Rousseau-Nepton L. et al., 2018, MNRAS, 477, 4152} ||
\footnotesize{Rousseau-Nepton L. et al., 2019, MNRAS, accepted, arXiv:1908.09017}  ||
\footnotesize{Sabbadin F., Minello S., Bianchini A., 1977, A\&A, 60, 147}

\end{document}